\documentclass[twocolumn]{emulateapj}
\usepackage{graphicx}

\begin{document}
%%%%%%%%%%%%%%%%%%%%%%%%%%%%%%%%%%%%%%%%%%%%%%
\shorttitle{FLUX ANOMALIES IN 1RXS J1131$-$1231}
\shortauthors{BLACKBURNE, POOLEY, \& RAPPAPORT}
\submitted{Accepted to ApJ}
%%%%%%%%%%%%%%%%%%%%%%%%%%%%%%%%%%%%%%%%%%%%%%
\title{X-Ray and Optical Flux Anomalies in the Quadruply Lensed QSO 1RXS J1131$-$1231\footnotemark[1]}

\author{J.\ A.\ Blackburne\altaffilmark{2,6}, D.\ Pooley\altaffilmark{3,4}, and S.\ Rappaport\altaffilmark{5,6}}

\footnotetext[1]{Based on observations obtained with the Clay 6.5 m telescope at the Las Campanas Magellan Observatories, and with the {\em Chandra X-Ray Observatory}.}
\altaffiltext{2}{MIT Kavli Institute for Astrophysics and Space Research, 77 Massachusetts Ave, Cambridge, MA 02139; {\tt jeffb@space.mit.edu}}
\altaffiltext{3}{Astronomy Department, University of California at Berkeley, 601 Campbell Hall, Berkeley, CA 94720; {\tt dave@astron.berkeley.edu}}
\altaffiltext{4}{Chandra Fellow}
\altaffiltext{5}{MIT Kavli Institute for Astrophysics and Space Research, 77 Massachusetts Ave, Cambridge, MA 02139; {\tt sar@mit.edu}}
\altaffiltext{6}{MIT Department of Physics}
%%%%%%%%%%%%%%%%%%%%%%%%%%%%%%%%%%%%%%%%%%%%%%

\begin{abstract}

Optical and X-ray observations of the quadruply imaged quasar 1RXS J1131$-$1231 show flux ratio anomalies among the images factors of $\sim$2 in the optical and $\sim$3--9 in X-rays. Temporal variability of the quasar seems an unlikely explanation for the discrepancies between the X-ray and optical flux ratio anomalies. The negative parity of the most affected image and the decreasing trend of the anomalies with wavelength suggest microlensing as a possible explanation; this would imply that the source of optical radiation in RXS J1131 is $\sim 10^4 R_g$ in size for a black hole mass of $\sim 10^8 M_{\odot}$. We also present evidence for different X-ray spectral hardness ratios among the four images.

\end{abstract}

\keywords{gravitational lensing --- quasars: individual (1RXS J1131$-$1231)}

\section{Introduction}
\label{sec:intro}

1RXS J1131$-$1231 was first cataloged as an X-ray source by {\em ROSAT}, and was subsequently discovered by \citet{sluse03} to be a rather spectacular quadruple gravitational lens. It is a quasar at a redshift of 0.658, lensed by an elliptical (but nearly spherical) galaxy at a redshift of 0.295 into four images arranged in what \citet{saha03} call a {\em long-axis quad}. In addition, a portion of the host galaxy is lensed into an optical Einstein ring. Quadruple quasar lenses such as this are useful for lens modelers because of the relatively large number of constraints they provide. The models yield the mass of the luminous plus dark matter responsible for the lensing, as well as more limited information on the radial profile of the matter distribution in the lensing galaxy.  In spite of these successes, it has proven difficult in many cases to match the flux ratios among the images predicted by lens models, to those actually observed \citep{mao98,metcalf02}.

Small-scale structure is often invoked to explain these flux ratio anomalies, either in the form of microlensing by stars \citep{chang79,schechter02} or {\em millilensing} by dark matter subhaloes with masses of $\sim 10^6 \, M_\odot$ \citep{mao98,dalal02}. Another potential source of flux anomalies is differential extinction in the lensing galaxy (where the four beams are furthest from each other spatially). However, no significant optical or X-ray extinction is expected in the outskirts of an early-type galaxy such as the one responsible for the lensing in RXS J1131.

In this letter we report on X-ray observations made with the {\em Chandra Observatory} at a single epoch, and optical observations made at Magellan at six epochs over the course of fourteen months. We find that the optical brightness of the source varied by no more than 0.3 magnitudes during this time.  The optical flux ratios among the four images are discrepant with the lens model by factors of $\sim$2, while the corresponding X-ray flux anomalies are factors of $\sim$3--9!  In addition, the A and B images appear to have significantly different X-ray spectral hardness ratios than do the C and D images.  We briefly discuss some possible explanations for the flux ratio anomalies and the different spectral hardness ratios.

\section{Observations}
\label{sec:obs}

\subsection{X-Ray Observations}
\label{sec:xrayobs}

RXS J1131 was observed for 10.0 ks on 2004 Apr 12 (ObsID 4814) with the Advanced CCD Imaging Spectrometer (ACIS) on the {\it Chandra X-ray Observatory}.  Each ACIS chip has $1024 \times 1024$ pixels and is 8\farcm3 on a side (with a pixel size of 0\farcs49).  The point spread function (PSF) is both energy-dependent and position-dependent.  Near the aimpoint, the half-power diameter is about 0\farcs8 at 1 keV, broadening to about 1\arcsec\ at 8 keV.  The data were taken in timed-exposure mode with an integration time of 3.14 s per frame, and the telescope aimpoint was on the back-side illuminated S3 chip. The data were telemetered to the ground in very faint mode. 

The data were downloaded from the {\it Chandra} archive, and data reduction was performed using the {\tt CIAO\,3.2.2} software provided by the {\it Chandra} X-ray Center\footnote{\url{http://asc.harvard.edu}}.  The data were reprocessed using the CALDB\,3.1.0 set of calibration files (gain maps, quantum efficiency, quantum efficiency uniformity, effective area) including a new bad pixel list made with the {\tt acis\_run\_hotpix} tool.  The reprocessing was done without including the pixel randomization that is added during standard processing; this omission slightly improves the point spread function.  The data were filtered using the standard {\it ASCA} grades and excluding both bad pixels and software-flagged cosmic ray events. Intervals of background flaring were searched for, but none were found.

\begin{figure}
\includegraphics[width=0.39\textwidth]{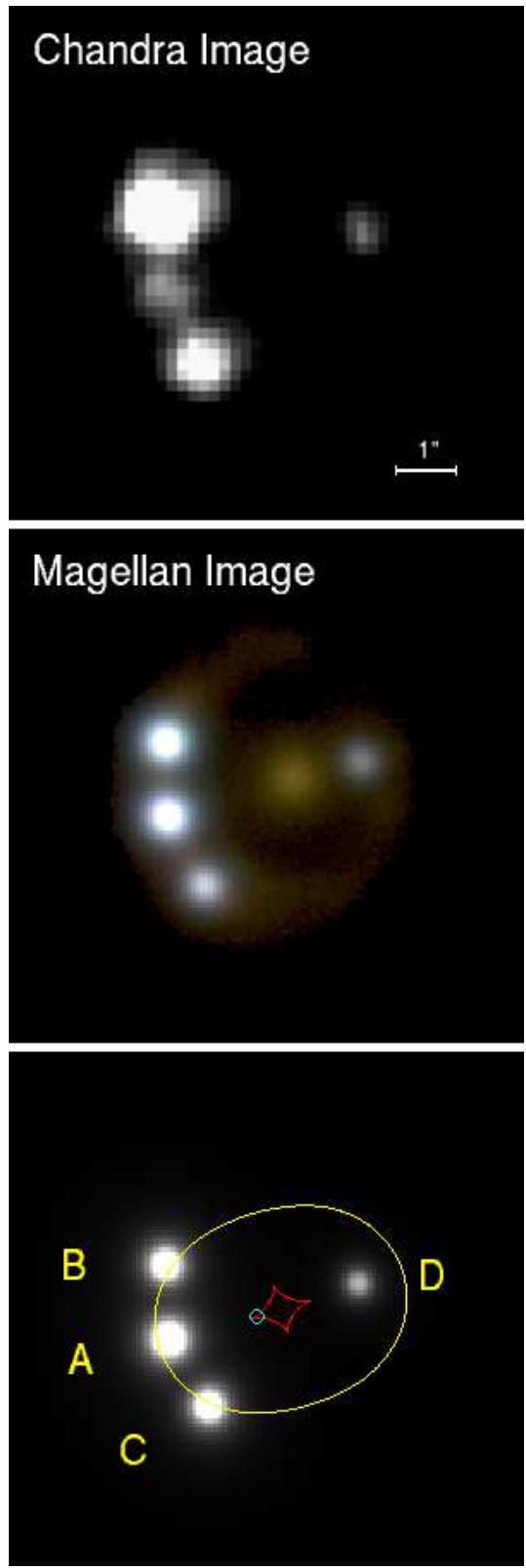}
\caption{\label{fig:pics}
{\it Chandra}, Magellan, and model images (top to bottom) of RXS J1131. The raw {\it Chandra} image was convolved with a Gaussian with a width $\sim$70\% the width of the {\it Chandra} PSF in order to produce a smoother appearance. The model image positions and brightnesses are from the model described in \S\ref{sec:model}, as are the predicted source and lens positions, marked with circle and diamond-shaped caustic, respectively. Also visible is the critical curve.}
\end{figure}

The IDL-based software package {\tt ACIS Extract} v3.79 \citep{broos02} was used for subsequent reduction and analysis.  An image of the X-ray data (see Figure~\ref{fig:pics}) was constructed by reprojecting the events around RXS J1131 in the 0.5--8 keV energy range using a spatial bin size of 0\farcs16\footnote{The satellite continuously dithers in a Lissajous pattern on the sky, requiring all images to be reprojected.  Standard processing produces an image with pixels that are 0\farcs492 on a side to match the physical CCD pixel size.  {\tt ACIS Extract} produces images with pixels matched to the size of those in the model PSFs.}.  Model PSFs were produced for the images using the {\tt CIAO} tool {\tt mkpsf} at energies of 0.277, 1.4967, 4.51, 6.4, and 8.6 keV.  The 1.4967 keV PSF was used in a maximum-likelihood reconstruction image of the data (10,000 iterations) in order to determine precise positions for each of the four lensed images.

Small apertures (about 0\farcs3 in radius) centered on these positions were used to extract counts and spectra, and the {\tt CIAO} tools {\tt mkacisrmf} and {\tt mkarf} were used to produce response files.  {\tt ACIS Extract} corrected the effective area response at each energy based on the fraction of the PSF enclosed by the extraction aperture at that energy, interpolating from the five model PSFs.  The apertures enclosed roughly 30\% of the PSF at 1.5 keV and roughly 25\% of the PSF at 6.4 keV.  These small apertures were desirable in order to reduce contamination from the other lens images, but image A still suffered some small contamination from images B and C.  To correct for image A being in the wings of the PSF of both B and C, five extraction regions were placed around B and five around C at the same radial distance as A.  The averages of each set of five regions were used for subtraction from A.  The contribution of the cosmic X-ray background in the lens extraction regions is negligible (roughly 0.005 counts).  

The spectrum of each lens image was fit in {\tt XSPEC\,12.2} \citep{arnaud96} with a powerlaw model absorbed by two components, one fixed at the Galactic column to RXS J1131 of $n_H = 3.64\times10^{20}$~cm$^{-2}$ \citep{dickey90} and one allowed to vary.  Acceptable fits were obtained for each image, with reduced $\chi^2$ of 0.37, 1.3, 0.59, and 1.8 for images A, B, C, and D, respectively.  For each of the images, the additional absorption components were consistent with zero, and the upper ends of the 1-$\sigma$ confidence intervals were 6.0, 1.4, 2.1, and $4.4\times10^{20}$~cm$^{-2}$, indicating similar (and small) absorbing columns for each image.  The powerlaw indices (with 1-$\sigma$ confidence intervals) were $1.21^{+0.27}_{-0.14}$, $1.23^{+0.08}_{-0.05}$, $1.63^{+0.17}_{-0.09}$, and $1.93^{+0.58}_{-0.29}$, indicating some intrinsic spectral differences (i.e., not due to absorption) among the lens images.  To further quantify these differences, the spectral hardness ratio ($SR$) was defined as the observed photon flux (photons~cm$^{-2}$~s$^{-1}$) in the 2--8 keV band to that in the 0.5--2 keV band (see Table~\ref{tab:lens}).  To characterize the intensity of each image, the unabsorbed power-law flux was integrated over 0.5--8 keV.  Table~\ref{tab:lens} lists this flux for image B and the flux ratio for the other images.

\begin{deluxetable}{c c c c}
\tablewidth{0pt}
\tablecaption{X-Ray and Optical Properties of RXS J1131
\label{tab:lens}}
\tablehead{
\colhead{} & \colhead{X-Ray} &
\colhead{Optical} & \colhead{Model} } %\\
\startdata
$F_B\tablenotemark{a}$ & $1.8 \times 10^{-13}$ & $7.5 \times 10^{-14}$ & 1 (+) \\
$F_A/F_B$ & $0.18 \pm 0.04$ &  $1.10 \pm 0.155$ & 1.703 (-)\\
$F_C/F_B$ & $0.27 \pm 0.03$ &  $0.47 \pm 0.063$ & 0.962 (+)\\
$F_D/F_B$ & $0.06 \pm 0.01$ &  $0.17 \pm 0.061$ & 0.113 (-)\\
$L_B\tablenotemark{b}$ & $3.2 \times 10^{44}$ &  $1.3 \times 10^{44}$
& \nodata \\[4pt]
\colrule
\\
SR(A)\tablenotemark{c} & $1.05 \pm 0.28$ & 1.10; 0.86 $\pm 0.13$ & \nodata \\
SR(B)\tablenotemark{c} & $1.14 \pm 0.09$ & 0.96; 0.94 $\pm 0.13$ & \nodata \\
SR(C)\tablenotemark{c} & $0.72 \pm 0.10$ & 0.96; 0.82 $\pm 0.13$ & \nodata \\
SR(D)\tablenotemark{c} & $0.58 \pm 0.16$ & 0.93; 0.77 $\pm 0.13$ & \nodata \\
\enddata
\tablenotetext{a}{Flux in units of ergs cm$^{-2}$ s$^{-1}$; corrected
for a magnification of 13.9, as determined from a model of the lens.}
\tablenotetext{b}{Luminosity in units of ergs s$^{-1}$ for a source
at \mbox{$z=0.658$} and a corresponding luminosity distance of 3840 Mpc. No
k-corrections have been made.}
\tablenotetext{c}{Spectral ratios in the X-ray and optical bands.
X-ray ratios are defined as the observed photon flux in the 2--8 keV band
to that in the 0.5--2 keV band. The first of the optical spectral
ratios is for the Sloan $g'$ band to $r'$ band; the second
is for the $r'$ to $i'$ bands. The optical spectral ratios
are given in linear flux units.}
\end{deluxetable}

\begin{deluxetable}{c c c c c}
\tablewidth{0pt}
\tablecaption{Optical Variability of RXS J1131
\label{tab:opt}}
\tablehead{
\colhead{Julian Date\tablenotemark{a}} & \colhead{Image A\tablenotemark{b}} &
\colhead{Image B\tablenotemark{b}} & \colhead{Image
C\tablenotemark{b}} & \colhead{Image D\tablenotemark{b}}} %\\
\startdata
2453055 & 17.69 & 17.70 & 18.49 & 19.71 \\
2453058 & 17.76 & 17.79 & 18.69 & 19.80 \\
2453135 & 17.56 & 17.52 & 18.51 & 19.95 \\
2453152 & 17.54 & 17.58 & 18.48 & 19.64 \\
2453376 & 17.85 & 18.16 & 18.82 & 19.67 \\
2453475 & 18.04 & 18.33 & 18.95 & 19.73 \\
\enddata
\tablenotetext{a}{Julian date of the Magellan observation.}
\tablenotetext{b}{Sloan AB $i'$ band magnitude, after correction for
reference stars in the same field of view.}
\end{deluxetable}

\subsection{Optical Observations}
\label{sec:optobs}

The lens was observed at six epochs over the course of fourteen months in 2004 and 2005 at the Magellan 6.5-meter Clay telescope at Las Campanas Observatory. The observations made use of the Raymond and Beverly Sackler Magellan Instant Camera (MagIC), a $2048 \times 2048$ direct imaging instrument with a plate scale of 0\farcs{}069 per pixel and a 2.4 arcminute field of view. One epoch of observations included imaging in three bands --- Sloan $g'$,$r'$, and $i'$; the others were limited to the $i'$ band only. The seeing on these nights varied from 0\farcs{}4 (on the night of multicolor observations) to 1\farcs{}0.

Three images from the night with the best seeing were used to produce the pseudocolor optical image shown in Figure \ref{fig:pics}. The $g'$ band was mapped to blue, while $r'$ was mapped to green and $i'$ to red. The color stretch has been matched to the square root of the flux to bring out the faint Einstein ring.

The {\tt DoPHOT} PSF-fitting photometry program was used to measure the positions and magnitudes of the four quasar components and the lensing galaxy, as well as five nearby reference stars. The presence of the Einstein ring, which has a red color and thus is especially strong in the $i'$ band, has resulted in some additional small uncertainties, both in astrometric and photometric measurements, which are difficult to quantify. We estimate the astrometric errors at 0\farcs{}01.

The standard stars 101 207 and RU 149F \citep{landolt92} were used to bring the multi-band observations to the AB magnitude system. The transformations from Johnson colors to Sloan AB colors were taken from \citet{fukugita96}. The colors thus obtained for the four quasar images are reported in Table \ref{tab:lens}, in linear flux units.

The five field stars were used to calibrate the $i'$ band photometry for all six epochs to the same magnitude system as the multi-band observations. After this normalization, the magnitudes of the 5 stars had rms fluctuations of between 0.02 and 0.07 magnitudes. The magnitudes of the four quasar components thus derived are presented in Table \ref{tab:opt}. Because of the effects of the Einstein ring, the uncertainties are larger for these than for the reference stars; we estimate them to be 0.15 magnitudes. Within this level of uncertainty, the data are almost consistent with a steady brightness over time, although there appears to be a slight dimming trend in the last three epochs.

\section{Discussion}
\label{sec:discuss}

\subsection{Modeling the Lens}
\label{sec:model}

Following the lead of \citet{sluse03}, we used Keeton's (2001) {\tt Lensmodel} software to model the lens as a singular isothermal sphere (SIS) with an external shear. Including the position of the source as well as the lens position and strength and the shear strength and direction, there were seven free parameters. We used the Magellan positions of the four lens components, which had uncertainties of 0\farcs{}01, for a total of eight constraints. We did not constrain the position of the lens, in order to allow for the possibility that mass may not strictly follow the light. The fit yielded a reduced $\chi^2$ of 1.1. The lensing mass was predicted to be 0\farcs{}14 southeast of the observed galaxy position, indicating that our model is not perfect; this is typical of such simple lens models. We find the lens strength to be 1\farcs{}78, and the shear to be 0.12 in a direction $73.3^{\rm o}$ west of north. These values are similar to those reported by \citet{sluse03} of 1\farcs{}82, 0.12, and $14.8^{\rm o}$ east of north\footnote{A $90^{\rm o}$ offset between the two position angles is due to differing sign conventions for the shear term.}.

The magnifications predicted by the best fit model are -23.7, 13.9, 13.4, and -1.58 for images A, B, C, and D, respectively, with signs indicating image parity. These appear as flux ratios in Table \ref{tab:lens}. The $F_C/F_B$ and $F_A/F_B$ ratios are low by factors of $\sim$2 in the optical, and by factors of $\sim$3--9, respectively, in X-rays. The model relative intensities were used to create a simulated image as it would appear through the Magellan telescope. This image is shown in Figure \ref{fig:pics}. It is clear that the predicted flux ratios are different from those observed.

\subsection{Genuine Optical Anomalies}
\label{sec:genuine}

The X-ray flux ratios clearly appear to be anomalous, but one may fairly wonder if another relatively simple lens model might fit the optical data better. \citet{keeton03} use the ``cusp relation'' (which predicts in a model-independent way that the flux of image A should be approximately the same as the sum of images B and C) to establish convincingly that a simple smooth model with an elliptical galaxy cannot explain the optical flux ratios in RXS J1131. A highly flattened model such as an edge-on disk might explain them, but the morphology of the galaxy and the round Einstein ring rule out this possibility.

Our own modeling efforts bore out this conclusion. We constrained the fluxes to equal the optical values and modeled the lens both as an isothermal ellipsoid with external shear, and as an isothermal sphere with another isothermal sphere off-center, to provide shear. These models did not fit nearly as well as our best fit model above, with the extra contribution to $\chi^2$ coming nearly exclusively from the flux constraints. We also tried loosening the constraints on the positions of the images. This did improve the flux fits somewhat, but caused the positions to be fit far from their observed values.

\subsection{Quasar Variability}
\label{sec:variability}

Another possibility for explaining the differences in the X-ray and optical flux ratio anomalies might be to invoke temporal variability in the intrinsic output of the quasar, since the observations in the two wavebands were made at different epochs -- though the X-ray observation was made about half way between the second and third of the optical observations.  We have shown directly that the optical flux did not undergo any major secular changes in the intensity during that year. RXS J1131 has a luminosity which is about midrange (on a log scale) for quasars. Therefore it {\em may} undergo substantial temporal variability in both its intensity and spectral slope \citep[see, e.g.,][]{green93}. However, a sustained ($\gtrsim 10^4$ s) change in intensity by a factor of $\sim$9 within a day (the time delay difference between images A and B) would be quite unusual (Green et al. 1993). Thus, it seems doubtful that temporal variability explains the principal flux ratio anomaly in this source.

\subsection{Anomalies Due to Substructure}
\label{sec:explain}

As is the case for most anomalous quadruply lensed quasars \citep{kochanek03}, the sense of the flux anomaly in RXS J1131 is to demagnify the brightest saddle point image (image A), and possibly to further magnify the brightest minimum (image B). This discrimination by image parity is expected for both micro- and millilensing (Schechter \& Wambsganss 2002; Kochanek et al. 2003). In addition, the fact that the anomaly in image A is more severe in X-rays, which originate from a smaller region than the optical light, supports microlensing rather than millilensing.

To help understand what substructure might do to the intensities of the optical and X-ray images, we estimate the ratio of the angular sizes of the emitting regions near the quasar black hole to the Einstein radius of a point object (e.g., a star) in the lensing galaxy \citep[see also][]{mortonson04}.  Objects orbiting near the central black hole at radius $r$ subtend a characteristic angle at the earth of:
\begin{equation}
\theta_s \simeq \frac{r}{D_S} = \left(\frac{r}{R_g}\right) \frac{GM_{BH}}{c^2D_S} ~~~,
\end{equation}
where $M_{BH}$ is the black hole mass, $R_g$ is the gravitational radius of the black hole, and $D_S$ is the angular-diameter distance to the source.  By comparison, the Einstein radius of a micro- or millilensing point mass is:
\begin{equation}
\theta_E = \left(\frac{4GM_{mL}D_{LS}}{D_LD_Sc^2}\right)^{1/2} ~~~,
\end{equation}
where $M_{mL}$ is the mass of the micro- or millilensing object, and $D_L$ and $D_{LS}$ are the lens and lens-to-source angular diameter distances, respectively.

We can define a dimensionless ratio of these quantities $\xi$, which is related to the degree to which micro- or millilensing can occur:
\begin{eqnarray}
\xi \equiv \frac{\theta_s}{\theta_E} = 3.5 \times 10^{-4} \left(\frac{r}{R_g}\right) \times ~~~~~~~~~~~~~~~~ \nonumber \\
\left(\frac{M_{BH}}{10^8M_\odot}\right)\left(\frac{M_{mL}}{M_\odot}\right)^{-1/2} \sqrt{\frac{D_L{\rm (Gpc)}}{D_SD_{LS}}}~~~.
\end{eqnarray}
For RXS J1131 the above expression reduces to 
\begin{equation}
\xi \simeq 3 \times 10^{-4} \left(\frac{r}{R_g}\right)\left(\frac{M_{BH}}{10^8M_\odot}\right)\left(\frac{M_{mL}}{M_\odot}\right)^{-1/2} ~~~,
\end{equation}
\begin{equation}
{\rm or},~~~\xi \simeq 3 \left(\frac{\beta}{0.01}\right)^{-2}\left(\frac{M_{BH}}{10^8M_\odot}\right)\left(\frac{M_{mL}}{M_\odot}\right)^{-1/2} ~~~,
\end{equation}
where $\beta$ is the characteristic speed of orbiting or free-fall objects around the black hole, and we have taken $D_S\simeq1400$, $D_L\simeq900$, and $D_{\rm LS}\simeq 865$, all in units of Mpc.

Thus, X-ray and optical continuum emission which is emitted by the accretion disk within several hundred $R_g$ of the black hole can be substantially microlensed (see eq. (4)). Any broad-line emission features (with $\beta \simeq 0.01$) could be only marginally microlensed (see eq. (5)). In contrast, any narrow-line emission region (with $\beta \simeq 10^{-3}$) would {\em not} be microlensed.  In this study, we are limited to X-ray and continuum optical emission, both of which should be about equally microlensed.

Therefore, the clear differential in the flux ratio anomalies between the optical and X-ray bands, factors of $\sim$2 in the former, and $\sim$3--9 in the latter, presents something of a puzzle (see \S\ref{sec:summary}).  If, on the other hand, the continuum optical emission originates farther from the center, possibly due to scattering of visible light or reprocessing of higher energy radiation, then the differential flux anomalies between X-ray and optical could be explained by microlensing.  In this case we can directly estimate the size of the optical emission region as $\sim 10^4~R_g$ (see eq. (4)) for a $\sim$$10^8~M_\odot$ black hole.

Finally, to determine if it is plausible to explain a factor of 9 demagnification using microlensing, we examined the microlensing simulations described by \citet{schechter02}. For a saddle-point image with a magnification of $\sim$20 such as image A, the probability of a demagnification a factor of 9 or greater ranges from virtually zero for a 100\% stellar local projected mass density to nearly 17\% for a mass density made of 10\% stars and 90\% smooth dark matter. We expect that at this distance from the galaxy's center, stars would make up about 15--30\% of the projected mass, and so conclude that it is possible for microlensing to explain the X-ray anomaly.

\section{Summary and Conclusions}
\label{sec:summary}

 We have analyzed optical and X-ray images of the quad lens 1RXS J1131$-$1231 and find anomalous flux ratios among the four images that are different in the optical than in the X-ray, with the more extreme anomalies being present in the X-ray band.  In particular, the ratio $F_A/F_B$ is a factor of $9.4\pm 1.7$ smaller in the X-ray band that is predicted from the model image.  The effects of microlensing in connection with anomalous flux ratios have been discussed extensively in the literature (see, e.g., Metcalf \& Zhao 2002; Mortonson et al. 2004).  Above, we discuss why we would nominally expect the microlensing of both the continuum optical and the X-ray images to be almost the same.  
 
If the flux ratio anomaly differences between the X-ray and optical are ultimately resolved via microlensing, then we can turn the argument around and infer the approximate dimensions of the optical emission region (see, e.g., Mortonson et al. 2004).  First, we define $f \equiv L_{x+opt}/L_{edd}$, where $f$ is the fraction of the Eddington limiting luminosity that the X-ray plus optical luminosity represents.  The parameter $f$ incorporates the fact that $L_{x+opt}~(\sim 5 \times 10^{44}$ ergs s$^{-1}$) is less than the bolometric luminosity which, in turn, is less than Eddington.  The mass of the black hole can then be written as $(M_{BH} \simeq 2 \times 10^6~M_\odot)/f$.  Equation (4) can then be recast as:
\begin{equation}
r_{opt} \simeq 10^4R_g \left(\frac{f}{0.02}\right)~~~,
\end{equation}
where $r_{opt}$ is the size of the optical emission region, and we have taken $\xi \simeq 3$ in order to weaken the microlensing significantly (see, e.g., Schechter \& Wambsganss 2002; Mortonson et al. 2004), and $M_{mL} \simeq 1~M_\odot$.

Perhaps even more intriguing than the flux ratio anomalies is the evidence we have found for different spectral hardness ratios in the X-ray band among the four images (see Table \ref{tab:lens}).  To test for the statistical significance of these different hardness ratios, we divided all the X-ray events from the region of the {\it Chandra} image displayed in Figure \ref{fig:pics} into two bands, 0.5--2 keV and 2--8 keV, and formed a single hardness ratio.  This turned out to be $SR = 0.97 \pm 0.038$. We then computed spectral hardness ratios for each of the four individual images (see Table \ref{tab:lens}).  Finally, we evaluated the difference of each of these spectral hardness ratios from the overall mean and divided by the statistical uncertainties in the ratio determinations.  We find that image B has an $SR$ value higher than the mean by 1.7$\sigma$, while images C and D both have lower $SR$ values than the mean by 2.4$\sigma$. Comparing $SR$ values among the individual images, A and B are consistent with having the same $SR$, as are C and D.  However, image B has an $SR$ value which differs from those of both C and D by 3.1$\sigma$. Our more detailed spectral analyses of the individual sources (see \S\ref{sec:xrayobs}) shows rather convincingly that there is very little absorbing column density, and therefore low-energy absorption cannot be the cause of the different X-ray spectral hardness ratios.  Since there are time delays among the four different images, one might possibly invoke temporal variability as an explanation. While this is possible, the C and D images have similar spectral hardness ratios, but a relatively long time delay of $\sim$80 days, while the B and C images have quite different hardness ratios and a relatively short time delay of only a few hours. So in order to explain the spectral ratio anomalies using temporal variability, one must invoke a hardening of the source in the 80 or so days between D and A, which would last at least a day to accommodate the spectral hardness of B, followed by a very rapid softening in time for C only a few hours later. An extensive monitoring campaign of RXS J1131 with the {\em Chandra Observatory} would be very valuable in terms of better understanding what temporal variability actually occurs in this source.

\acknowledgements 

We thank Paul Schechter, Alan Levine, and Adam Bolton for extremely helpful discussions, Joachim Wambsganss for the use of his microlensing simulations, and the anonymous referee for thoughtful suggestions. SR acknowledges support from NASA Chandra Grant NAG5-TM5-6003X.  DP gratefully acknowledges support provided by NASA through Chandra Postdoctoral Fellowship grant number PF4-50035 awarded by the Chandra X-ray Center, which is operated by the Smithsonian Astrophysical Observatory for NASA under contract NAS8-03060. JB acknowledges support from NSF Grant AST-0206010.

\end{document}